\documentclass{PoS}

\usepackage{graphicx}
\usepackage{subfig}
\usepackage[font=footnotesize]{caption}

\title{Scintillation light production, propagation and detection in the 4-ton dual-phase LAr-TPC demonstrator (data analysis and simulations)}

\ShortTitle{Scintillation light in the 4-ton dual-phase demonstrator}

\author{\speaker{Chiara Lastoria}\\ \thanks{on behalf of the DUNE Collaboration}\\
        CIEMAT, Centro de Investigaciones Energéticas, Medioambientales y Tecnológicas - Madrid (Spain)\\
        E-mail: \email{chiara.filomena.lastoria@cern.ch}}

\abstract{The Deep Underground Neutrino Experiment (DUNE) Far Detector (FD) will be formed by four 10-kton Liquid Argon (LAr) Time Projection Chambers (TPC) using both single and dual-phase technology. 
The dual-phase technology foreseen the charge amplification in the gas phase before the signal collection and is following a staged approach to demonstrate its feasibility at the DUNE FD scale. In 2017, a 4-ton demonstrator of 3x1x1 m$^3$ volume was exposed to cosmic muons and demonstrated expected performance in terms of charge extraction and light collection. A bigger prototype (ProtoDUNE-DP), with an active volume of 6x6x6 m$^3$, is currently under commissioning at CERN. \\
The photon detection system in these detectors is crucial to provide the trigger signal giving an absolute time reference for the charge acquisition system of rare non-beam events, and to provide complementary calorimetry. An overview of the analysis of the light collected in the 4-ton demonstrator
has been presented. These prototypes confirmed the performance of the light detection system to provide trigger based on the scintillation light signal, to characterize the LAr response to the crossing muons and to monitor the LAr purity. The analyzed data are compared with MC simulations to improve the values of less understood LAr optical parameters such as the Rayleigh scattering length.}

\FullConference{XXIX International Symposium on Lepton Photon Interactions at High Energies - LeptonPhoton2019\\
		August 5-10, 2019\\
		Toronto, Canada}

\begin{document}

\section{Introduction}

DUNE \cite{DUNE_IDR} is a new generation long-baseline neutrino detector pursuing the goals of measuring neutrino oscillations, investigating the presence of CP-violation and performing neutrino astrophysics studies and nucleon decay searches. To accomplish these goals, DUNE aims to build a far detector, 1300 km far away from the beam production, made by 4 modules filled by 10 kton of LAr both in the single or in the dual-phase (DP) configuration. 

In addition to the drift of the electrons produced by the interaction of charged particles crossing the LAr volume, the DP configuration provides the amplification of these electrons from their extraction in the gas argon (GAr) phase, allowing longer drift distances. The charge signal, collected in the anode plane, will offer an excellent 3D track imaging and an optimized signal-to-noise ratio. Complementary and additional information to the charge signal can be obtained from the analysis of the light signals produced by the crossing particles in LAr (prompt scintillation light, S1) and in GAr (electro-luminescence light, S2). Since the luminescence light produced in Ar volume is produced at $\sim$128 nm, a wavelength shifter is needed to make possible the light detection by photomultipliers tubes (PMTs).

\section{Description of the detector, light detection system and trigger}

The DP technology operation at large scale will be demonstrated through the performance of two intermediate prototypes built at CERN. The 4-ton demonstrator \cite{4TON} was the smallest one, with an active volume of 3x1x1 m$^3$ filled with 4.2 tons of LAr; it was exposed to cosmic muons 
in 2017. The bigger ProtoDUNE-DP detector, with an active volume of 6x6x6 m$^3$ and filled with $\sim$ 300 tons of LAr, is currently under commissioning. 
The operation of the 4-ton demonstrator had an important role in the DP prototypes scaling since it demonstrated the feasibility of such technology at the ton scale. It showed the possibility to extract the drifted electrons over a 3 m$^2$ surface, to multiply them using 50x50 cm$^2$ 
LEM units and to read them out on two collection planes with strips with a maximum length of 3 m.

\begin{figure}[h!ptb]
  \centering 
  \subfloat{\includegraphics[width=0.35\linewidth, height=3.cm ]{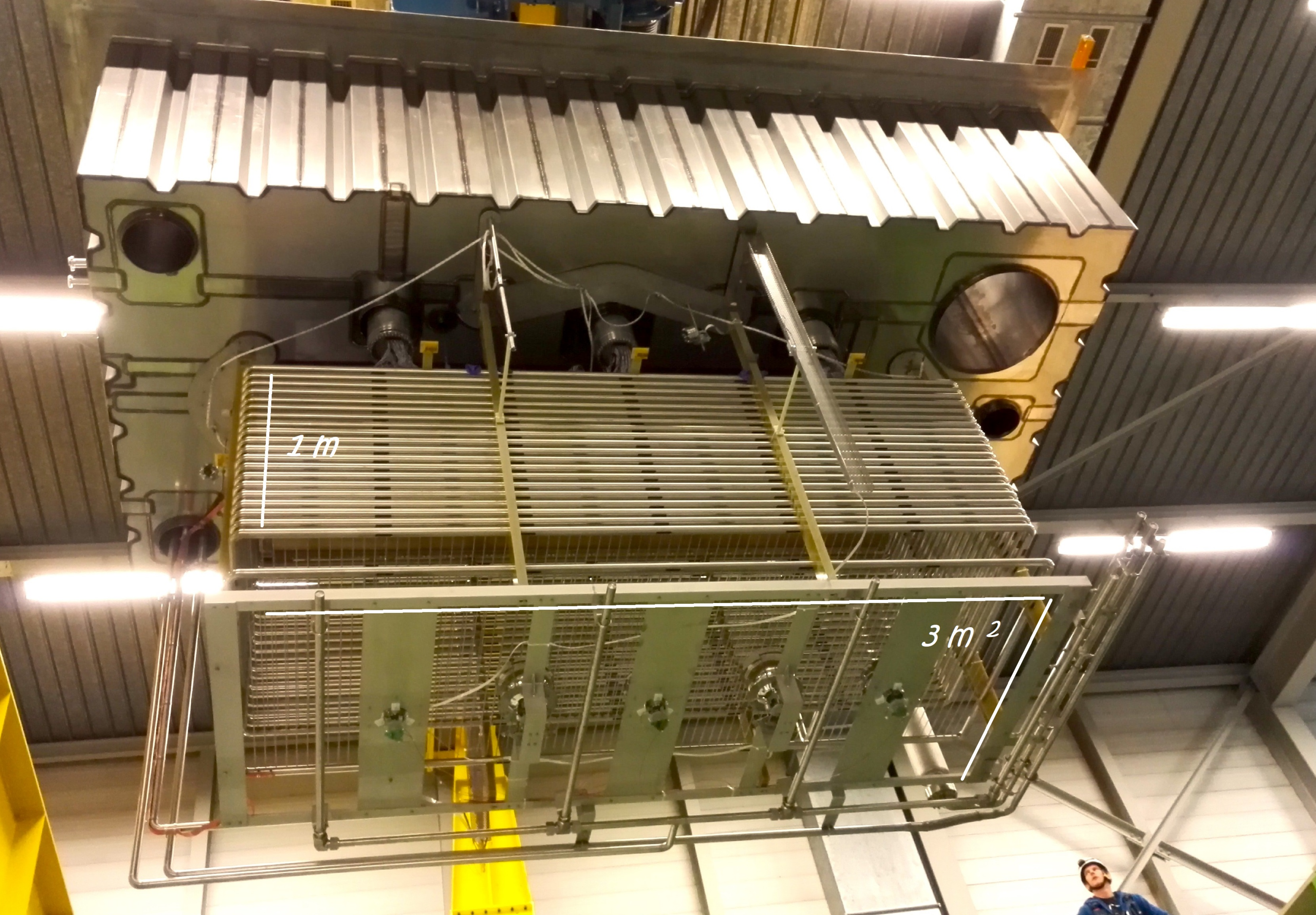}} 
  \hspace{0.5cm}
  \subfloat{\includegraphics[width=0.35\linewidth, height=3.cm ]{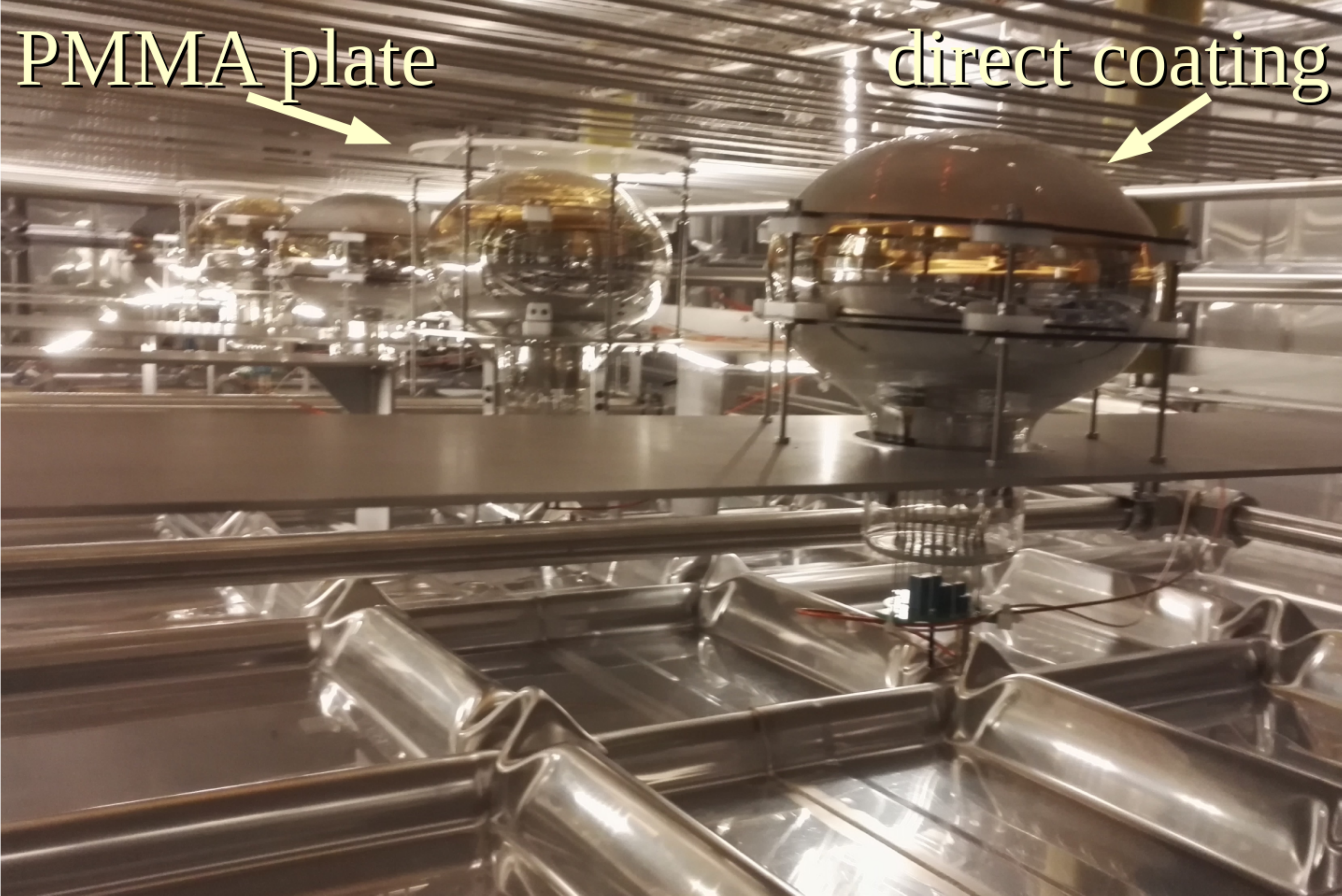}} 
  \caption{   Bottom view of 4-ton detector \textbf{(left)} and light detection system \textbf{(right)}. Two different configurations have been installed for the wavelength shifter, a direct TPB coating applied over the PMT surface or over two PMMA plates positioned above the PMT surface.\label{fig:311pictures}
  }
\end{figure}

The light detection system demonstrated it can provide the $\mathrm{t_{0}}$ time for the charge DAQ, so that it could be used as self trigger for the cosmic muons. To accomplish these requirements and to obtain complementary calorimetric information from the scintillation and electro-luminescence light signals, the light detection system used five eight inches cryogenic R5912-02Mod PMTs 
\cite{PMTs_ref} positioned underneath the cathode. To find the optimal configuration to be implemented in the DUNE DP FD module, the light detection system tested different PMT configurations in terms of base polarity (positive and negative bases) and wavelength shifter. In the demonstrator, Tetraphenyl-butadiene (TPB) was used as wavelength shifter in two different configurations, a direct coating applied over the PMT surface or over two Poly-methyl methacrylate (PMMA) plates above the PMT surface. Figure \ref{fig:311pictures} shows the 4-ton demonstrator detector (left) and the five PMTs with the two TPB coating options (right). 

As an alternative to the PMT self trigger based on a five fold coincidence of the prompt scintillation light signal, another trigger system was based on two pairs of Cosmic Rays Tagger (CRT) panels positioned outside the detector. The CRT trigger allowed direct track reconstruction, even in absence of drift field. Due to the geometrical acceptance of the CRT panels the trigger rate was around $\sim$ 0.3 Hz. On the other hand, even if the PMT trigger does not provide track information, this option increased the trigger rate up to $\sim$ 3 Hz.

\section{Light signal characterization in the 4-ton demonstrator}

The light analysis of the data collected characterize the S1 and S2 light signals. Since the knowledge of light propagation in the LAr is not completely understood, the combination of data analysis with the MC simulation is crucial. 
The light propagation is simulated with a Geant4-based software and the events are generated based on the CRT geometry. Other information related with the light propagation such as the Rayleigh scattering and the absorption lengths are introduced as parameters in the simulation. 

\begin{figure}[h!ptb]
  \centering 
  \subfloat{\includegraphics[width=0.33\linewidth, height=3.3cm ]{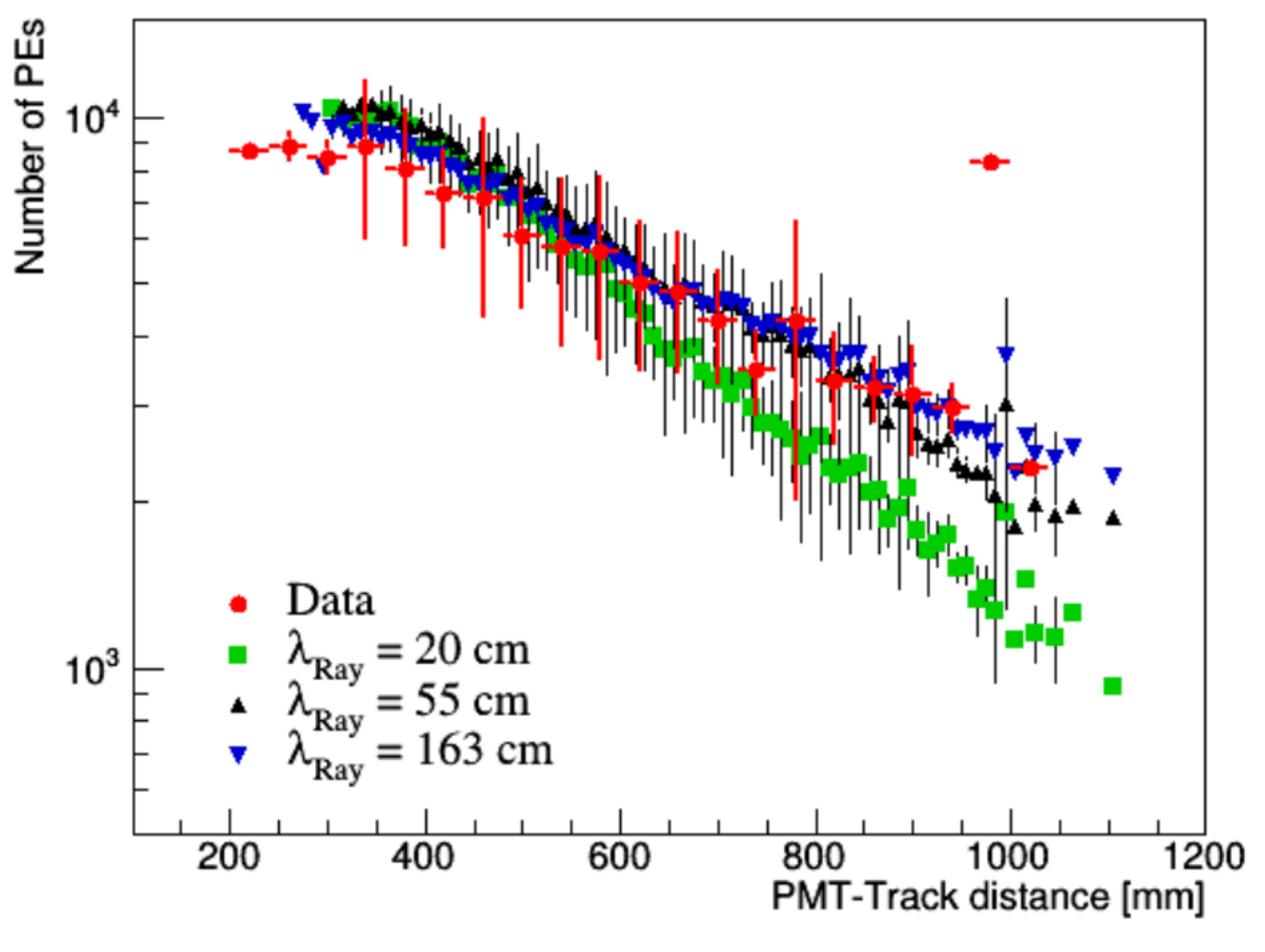}} 
  \hspace{0.1cm}
  \subfloat{\includegraphics[width=0.33\linewidth, height=3.3cm ]{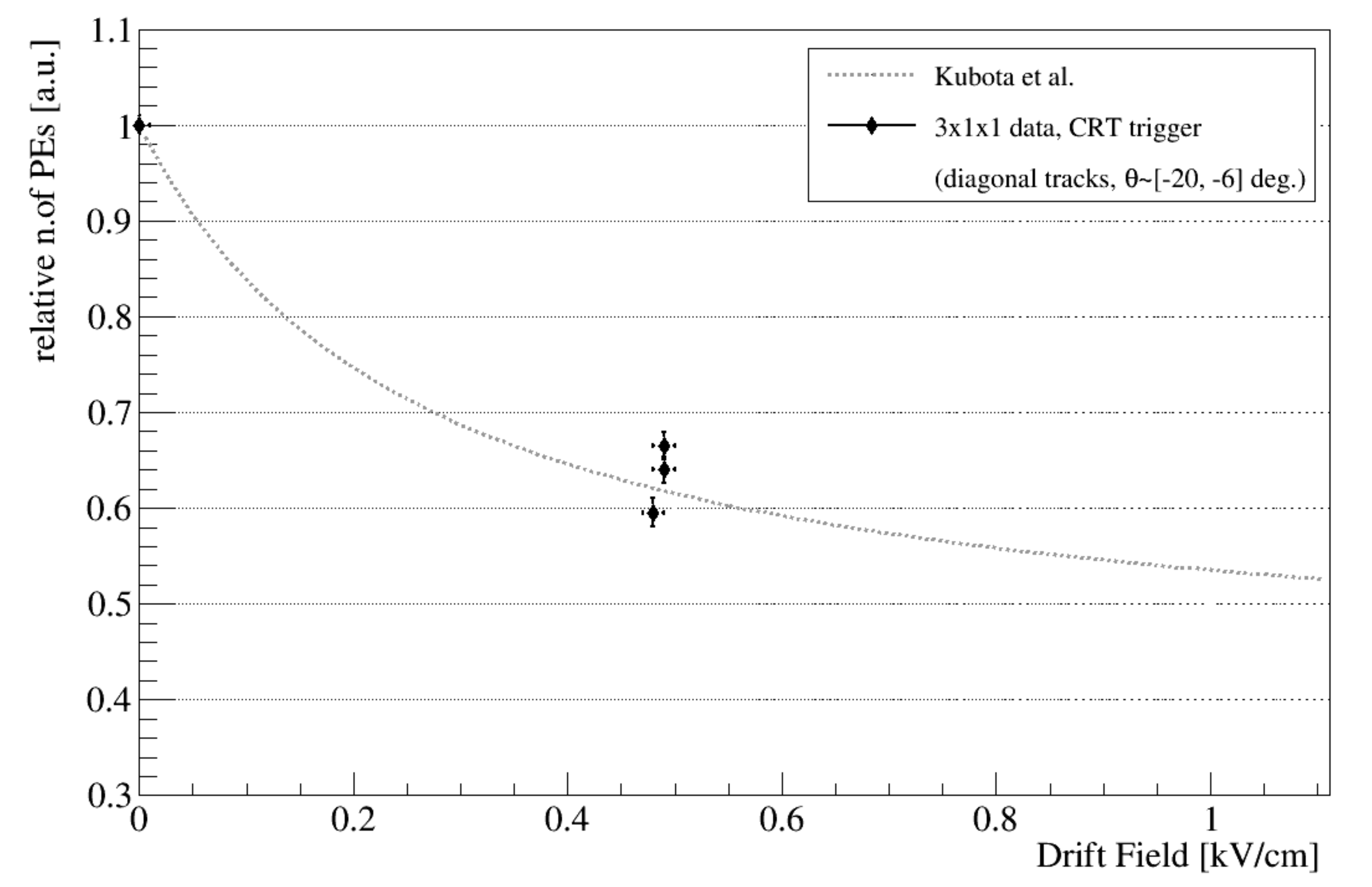}} 
  \hspace{0.1cm}
  \subfloat{\includegraphics[width=0.3\linewidth,  height=3.3cm ]{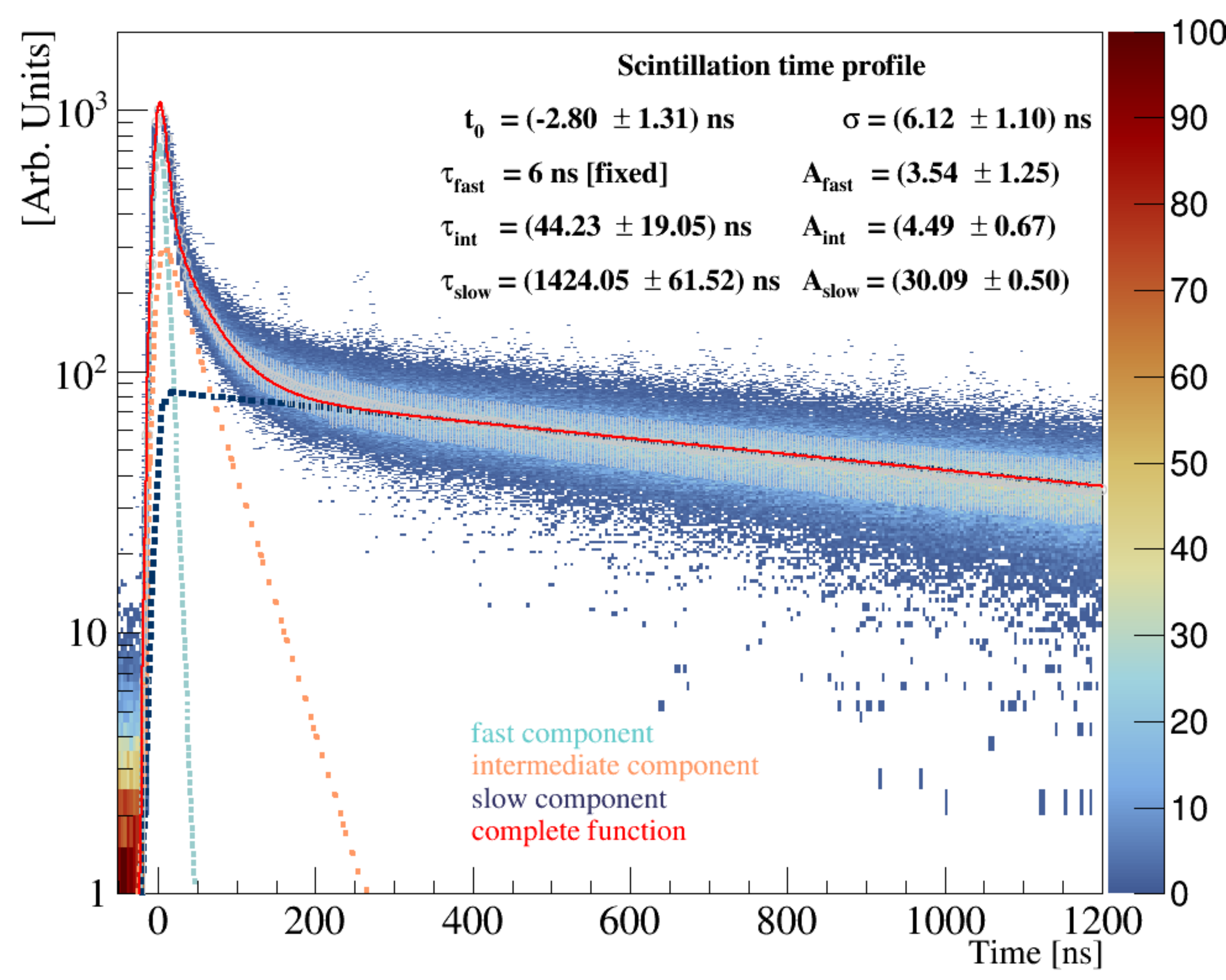}} 
  \hspace{0.1cm}
  \subfloat{\includegraphics[width=0.75\linewidth, height=3.3cm ]{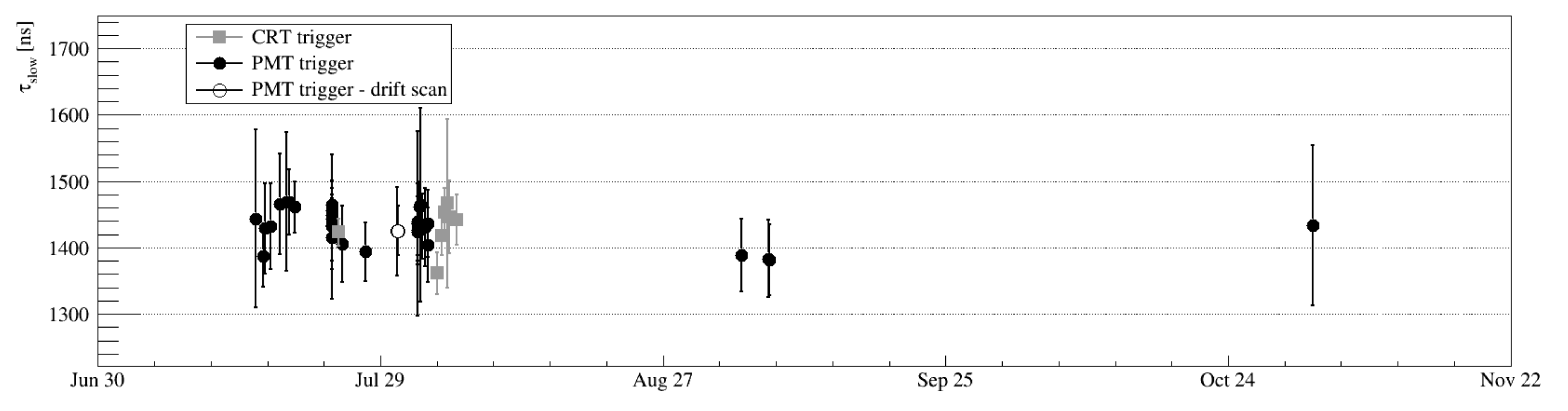}} 
  \caption{Characterization of the scintillation light produced in the 4-ton demonstrator. Data-MC comparison to estimate the Rayleigh scattering length \textbf{(top-left)}. Measurement of LAr recombination factor \textbf{(top-center)}. Example of the function used to fit the scintillation time profile \textbf{(top-right)}. Monitoring of the LAr purity through the $\tau_{slow}$ value obtained from the fit \textbf{(bottom)}.\label{fig:311_results}}
\end{figure}

An example of the importance of data-MC comparison is shown in top-left of Fig.\ref{fig:311_results}, the amount of collected light by each PMT is shown as a function of the track-PMT distance. Three MC simulations have been produced corresponding to three different Rayleigh scattering lengths (20cm, 55cm, 163cm) and the best agreement between data and MC is found for a Rayleigh scattering length between 55cm and 163cm.
In the top-center of the figure shows the measurement of LAr recombination factor considering the dependence of the light yield as a function of the drift field.
A more detailed characterization of the S1 signal has been obtained from the fit of the scintillation time profile that is shown in the top-right of the figure. The function used for the fit is the convolution of a gaussian function, needed to model the detector response, with three exponentials. Two exponential contributions are expected to fit the \emph{fast} and the \emph{slow} components corresponding to the de-excitations from the singlet or triplet levels LAr atoms; an additional \emph{intermediate} component is needed to fit the transition region between the previous two components. The sensitivity of the slow component to the possible presence of impurities affecting the scintillation light (mainly $\mathrm{O}_2$, $\mathrm{N}_2$ and $\mathrm{H}_2\mathrm{O}$) \cite{Acciarri} controls the LAr purity by the monitoring of the $\tau_{slow}$ value in all the runs collected in absence of drift field, as shown in the bottom part of Fig.\ref{fig:311_results}. The LAr purity corresponds to very small ($\lesssim$ ppm level) and stable amount of impurities since the $\tau_{\mathrm{slow}}$ remained, on average, slightly higher than 1.4$\mu s$.

A dedicated algorithm has been developed to identify the S2 signal produced when the electrons from ionization are drifted, extracted and amplified to the GAr phase. For each S2 signal the arrival time, the duration and charge can be retrieved. Due to the correlation between time difference $\Delta T_{\mathrm{S2-S1}}$ and the track geometry, a global measurement of the drift velocity can be performed obtaining (1.47 $\pm$ 0.12) mm/$\mu$s for the electrons drifted at 0.5 kV/cm. The characterization of the S2 signal and the relative MC simulation are being completed and updated results will be presented.

\section{Conclusions and Prospects}
The operation of the 4-ton demonstrator did the ground-breaking work for the ProtoDUNE-DP detector, confirming the scalability of the DP technology to the foreseen DUNE dual-phase FD module. The light detection system installed in the demonstrator showed the possibility to develop a trigger system based on the prompt scintillation light signal to detect cosmic muons, record the $\mathrm{t_0}$ time of the triggered event and to transmit this information to the charge DAQ system. The analysis of the S1 and S2 light signals allowed a better understanding of the LAr response to the crossing muons at the ton scale.

A paper is in preparation to describe the full scintillation light analyses. Pursuing and improving these analyses in the ProtoDUNE-DP detector are one of the goals of the experiment now in the commissioning phase.


\begin{thebibliography}{99}
\bibitem{DUNE_IDR} DUNE Collaboration, \emph{"The DUNE Far Detector Interim Design Report Volume 1: Physics, Technology and Strategies}, arXiv:1807.10334
\bibitem{4TON} B. Aimard, et al., \emph{"A 4 tonne demonstrator for large-scale dual-phase liquid argon time projection chambers"}, arXiv:1806.03317
\bibitem{PMTs_ref} Belver et al., \emph{"Cryogenic R5912-20Mod Photomultiplier Tube Characterization for the ProtoDUNE Dual Phase Detector"}, JINST \textbf{13} T10006
\bibitem{Acciarri} R. Acciari et al., \emph{"Effects of Nitrogen and Oxygen contaminations in liquid Argon"}, Nucl. Instrum. Methods Phys. Res. A \textbf{607} (2009) 169-172
\end{thebibliography}
\end{document}